\documentclass[useAMS]{mn2e}
\usepackage{graphicx}
\usepackage{amsmath}
\usepackage{dcolumn}% Align Table columns on decimal point
\def \aj {AJ}
\def \mnras {MNRAS}
\def \apj {ApJ}
\def \apjs {ApJS}

\def \spie {SPIE}

\title[Selecting Quasar Candidates by a SVM Classification System]{Selecting Quasar Candidates by a SVM Classification System}
\author[Nan-bo Peng, Yan-xia Zhang, Yong-heng Zhao and Xue-bing Wu]
{Nanbo Peng$^{1,2}$, Yanxia Zhang$^{1}\thanks{E-mail:
nbpeng@bao.ac.cn
(NP),zyx@bao.ac.cn (YZ)}$, Yongheng Zhao$^{1}$ and Xuebing Wu$^{3}$\\
$^{1}$Key Laboratory of Optical Astronomy, National Astronomical Observatories, Chinese Academy of Sciences 100012, Beijing, P.R.China\\
$^{2}$Graduate University of Chinese Academy of Sciences 100049, Beijing, P.R.China \\
$^{3}$Department of Astronomy, Peking University 100871, Beijing, P.R.China}
\begin{document}

\date{Accepted . Received 2011 }

\pagerange{\pageref{firstpage}--\pageref{lastpage}} \pubyear{2002}

\maketitle

\label{firstpage}

\begin{abstract}
We develop and demonstrate a classification system constituted by
several Support Vector Machines (SVM) classifiers, which can be
applied to select quasar candidates from large sky survey projects,
such as SDSS, UKIDSS, GALEX. How to construct this SVM
classification system is presented in detail. When the SVM
classification system works on the test set to predict quasar
candidates, it acquires the efficiency of 93.21\% and the
completeness of 97.49\%. In order to further prove the reliability and
feasibility of this system, two chunks are randomly chosen to
compare its performance with that of the XDQSO method used for
SDSS-III's BOSS. The experimental results show that the high faction
of overlap exists between the quasar candidates selected by this
system and those extracted by the XDQSO technique in the dereddened
i-band magnitude range between 17.75 and 22.45, especially in the
interval of dereddened i-band magnitude $<$ 20.0. In the two test
areas, 57.38\% and 87.15\% of the quasar candidates predicted by the
system are also targeted by the XDQSO method. Similarly, the
prediction of subcategories of quasars according to redshift
achieves a high level of overlap with these two approaches.
Depending on the effectiveness of this system, the SVM
classification system can be used to create the input catalog of
quasars for the GuoShouJing Telescope (LAMOST) or other
spectroscopic sky survey projects. In order to get higher confidence of quasar candidates, cross-result from the candidates selected by this SVM system with that by XDQSO method is applicable.

\end{abstract}


\section{Introduction}
\label{sec:intro}

Over the years, the volume of astronomical data at different
wavebands grows dramatically with large space-based and ground-based telescopes surveying the
sky, such as SDSS, 2MASS, NVSS, FIRST and 2dF.
How to preselect scientific targets from the enormous amount of observed data is a
significant and challenging issue.
In another words, how to extract knowledge from a huge volume of data
by automated methods is an important task for astronomers.
In the next decade,
the ongoing or planned multiband photometric survey projects, for instance,
the Large Synoptic Survey Telescope (LSST; Tyson 2002),
the Visible and Infrared Survey Telescope for Astronomy (VISTA; McPherson et~al. 2006),
and the Panoramic Survey Telescope and Rapid Response System (Pan-STARRS; Kaiser et~al. 2002) will bring more serious challenges for astronomers.

Ball and Brunner (2009) reviewed the current state of data mining and machine learning in astronomy. Borne (2009) also described the
application of data mining algorithms to research problems in astronomy. A lot of Data Mining (DM) algorithms have been applied to find
quasar candidates in astronomy. Traditional quasar selection relies on cutoff in a two-dimensional color space
although most modern surveys are done in several bandpasses. Traditional methods can't make use of the provided information from the high
dimensional space. Otherwise, the DM approaches utilize the features as many as possible. In general, DM methods for quasar candidate
selection can be divided into two types: supervised and unsupervised learning.
Most methods used in this domain of astronomy belong to supervised learning.
Abraham et~al. (2010) used a Difference Boosting Neural Network (DBNN) classifier
which is a bayesian supervised learning algorithm to make a catalogue of quasar candidates from the Sloan Digital Sky Survey Seventh Data
Release (SDSS DR7; Abazajian et~al. 2009). Carballo et~al. (2008) obtained a sample set of redshift $\ge$ 3.6 radio quasi-stellar objects using Neural Network (NN).
Quasar candidate detection also can be achieved by using unsupervised clustering algorithms in color spaces, such as the Probabilistic
Principal Surface(PPS) algorithm (D'Abrusco, Longo \& Walton 2009). The most representative work could be the series of work completed by
the SDSS team until now, especially for the SDSS-III Baryon Oscillation Spectroscopic Survey (BOSS; Schlegel et~al. 2007; Eisenstein et~al. 2011).
Ross et~al. (2011) gave a flowchart for the BOSS quasar target selection and exploited several methods including an Extreme-Deconvolution method
(XDQSO; Bovy et~al. 2011), a Kernel Density Esitimator (KDE; Richards et~al. 2004, 2009), a Likelihood method which likes KDE
(Kirkpatrick et~al. 2011) and a Neural Network method (NN; Y\`eche et~al. 2010) in this flowchart. After several times comparisons
of the efficiency of quasar selection methods, XDQSO was declared to be CORE for the rest of the BOSS quasar survey.

Support Vector Machines (SVM) is a supervised learning method and it
can produce a non-probabilistic binary linear classifier given a set
of training examples each of which is labeled as one of two
categories. SVM provides a good out-of-sample generalization and can
be robust, even when the training sample has some bias. This
distinguishing feature of SVM attracts many astronomers to use it
for selecting quasar candidates. Zhang \& Zhao (2003) applied two
classification algorithms, Support Vector Machines (SVM) and
Learning Vector Quantization (LVQ), to study the distribution of
various astronomical sources in the multidimensional parameter
space. Zhang \& Zhao (2004) demonstrated that SVM can show better
performance than Learning Vector Quantization (LVQ) and Single-Layer
Perceptron (SLP) when preselecting AGN candidates. Gao et~al. (2008)
compared the performance of SVM with K-Dimensional Tree (KD-Tree) to
separate quasars from stars and provide a good parameter combination
of magnitudes and colors for SVM. Bailer-Jones et~al. (2008)
developed and demonstrated a probabilistic method for classifying
quasars in surveys, named the Discrete Source Classifier (DSC) which
is a supervised classifier based on SVM. Kim et~al. (2011) presented
how to use SVM to do a variability selection for quasars on a set of
extracted time series features including period, amplitude, color
and autocorrelation value. In this work, we focus on constructing a
kind of classification system based SVM and use it to select quasar
candidates for the Chinese GuoShouJing Telescope (LAMOST).

This paper is organized as follows. Section~\ref{sec:data} describes the characteristics of data used in this experiment in detail.
In Section~\ref{sec:method}, we presents the brief of SVM, and how to use it to construct a SVM classification system. Section~\ref{sec:pef}
demonstrates the performance of this method for separating quasars from stars in a test set. The comparison of this system with the XDQSO method
for classifying quasars and stars will be discussed in Section~\ref{sec:qso_sel}. In Section~\ref{sec:conclusion}, we give the conclusion about our method
and what should be improved in the future work.

\section{The Data}
\label{sec:data}
% three parts
% DR7 specphotoAll
% DR8 plus specphotoAll
% DR8 photometric sources

The Sloan Digital Sky Survey (SDSS) is one of the most ambitious and influential surveys in the history of astronomy (York et~al. 2000). The SDSS used
a dedicated 2.5-meter telescope at Apache Point Observatory, New Mexico, equipped with two powerful special-purpose
instruments. The 120-megapixel camera imaged 1.5 square degrees of sky at a time, about eight times the area of the full moon. Over eight years of
operations (SDSS-I, 2000-2005; SDSS-II, 2005-2008), it obtained deep, multi-color images covering more than a quarter of the sky and
created 3-dimensional maps containing more than 930,000 galaxies and more than 120,000 quasars. Meanwhile, SDSS is continuing with the Third Sloan Digital Sky Survey (SDSS-III), a program of four new surveys using SDSS facilities.
SDSS-III began observations in July 2008 and released its first public data as Data Release 8 to emphasize its continuity with previous
SDSS releases. SDSS-III will continue operating and releasing data through 2014. SDSS-II carried out three distinct surveys: the Sloan
Legacy Survey, SEGUE (the Sloan Extension for Galactic Understanding and Exploration), the Sloan Supernova Survey. SDSS-III builds on
the legacy of the SDSS and SDSS-II to generate high-quality scientific data and to make important new discoveries. SDSS-III has been
designed to maximize understanding of three scientific themes: Dark energy and cosmological parameters, the structure, dynamics, and
chemical evolution of the Milky Way, the architecture of planetary systems.

The creation of a good classifier depends on a complete and representive training
sample. Therefore careful preparation of training sample is of great
importance. In this
specific problem, we just care about separating quasars from stars
and thus exclude extended sources (GALAXY). The training sets and test sets used in this method are produced from four data sets Quasar Catalogue V (Schneider et~al. 2010), SDSS DR7 (Abazajian et~at. 2009), SDSS DR8(Aihara et~al. 2011) and SDSS-XDQSO (Bovy et~al. 2011). In this section, we simply introduce these four data sets and how to use them to construct the training set for each SVM classifier in detail will be discussed in Section 3.2.

Based upon the SDSS DR7, quasar Catalogue V contains 105,783 (LowZ\_No 88201, MedZ\_No 14063, HighZ\_No 3519) spectroscopically confirmed quasars and represents the
conclusion of the SDSS-I and SDSS-II quasar survey.
In the following, LowZ\_QSO, MedZ\_QSO and HighZ\_QSO are short for
low-redshift quasars, medium-redshift quasars and high-redshift
quasars, respectively. According to the paper (Bovy et~al. 2011),
the definition of low-redshift,
medium-redshift and high-redshift corresponds to $z < 2.2$, $2.2
\leq z \leq 3.5$ and $z> 3.5$, separately.
For the several training sets in our SVM classification system, nine tenths of quasars (95,202 quasars including 79,421 LowZ\_QSO, 12,610 MedZ\_QSO and 3,171 HighZ\_QSO) of this catalogue are randomly sampled to construct them and the remaining one tenth of quasars (10,581 quasars including 8,780 LowZ\_QSO, 1,453 MedZ\_QSO and 348 HighZ\_QSO) will be used as test samples of quasars.

The training sample of stars consists of three parts. The first part
is from the spectral confirmed stars of SDSS DR8, the second part
comes from the unidentified pointed sources with {\it psfMag\_i} $<
17.75$ in the subarea of Stripe-82, the third part is made up of the
pointed sources with deredened $i$-band magnitude between 17.75 and
22.45 mag in the same subarea of Stripe-82 removing those predicted
by SDSS-XDQSO as quasars (the probability of quasars $> 0.5$). The
detailed information about the three parts is described as follows.

The spectral confirmed stars used in training sets are produced from
{\it SpecPhotoAll} Table in SDSS DR8 using the SQL interface to
Catalog Archive Server (CAS) mainly following the criteria described
in Section 3.2.1 of Richards et~al. 2002. Some records in the {\it
SpecPhotoAll} Table of SDSS DR8 should be removed because the sky
survey plan makes some sources to be duplicately observed several
times and some spectroscopically identified objects don't have
photometric corresponding sources. We set the attribute {\it class}
$= STAR$ which means this record is a stellar object, {\it
sciencePrimary} $= 1$ which represents the best version of spectrum
at this location, {\it Mode} $= 1$ which denotes this record with
the best photometric data and {\it zWarrning} $= 0$ to ensure the
{\it subclass} of STAR more reliable. The records with fatal errors
are excluded using {\it flags} such as BRIGHT, SATURATED, EDGE and
BLENDED. We also reject the objects whose magnitude errors are
larger than 0.2 in all five optical bands. In addition, a very few
records with the same {\it objID } are weeded out. Finally, we get a
catalog of 480,878 spectral confirmed stars from {\it SpecPhotoAll}
Table of SDSS DR8 and randomly sampled out two thirds (No. 320584)
of them for training and the rest (No. 160,294) of them for test.

The sample of photometric stars without spectra is constructed from
the {\it PhotoObjAll} table in SDSS DR8 using {\it mode} $= 1$, {\it
type} $= 6$, {\it specObjID} $= 0$ and {\it psfMag\_i} $< 17.75$.
Since SDSS Stripe-82 (Abazajian et~al. 2009) has been observed many
times, the data from this area are reliable. The point sources in
this area with the {\it psfMag\_i} $< 17.75$ can rarely be quasars,
so these photometric sources are regarded as stars. Actually, we
select a subarea which covers 150 deg$^2$ (-$30^\circ <
\alpha_{\mathrm{J2000}} <+30^\circ$ and -$1^\circ.25 <
\delta_{\mathrm{J2000}} < +1^\circ.25$) and this area was also
chosen by SDSS-XDQSO (Bovy et~al. 2011). Consequently, these are
115,010 photometric stars in this subarea of Stripe-82 with the {\it
psfMag\_i} $< 17.75$.

SDSS-XDQSO method is one of methods which serve SDSS-III for
targeting quasars. It uses the extreme-deconvolution method to
estimate the underlying density of stars and quasars in flux space
and then it convolves this density with flux uncertainties when
evaluating the probability that an unknown object is a quasar. In
recent blind tests of SDSS-III, it demonstrates a good performance
to the faint objects. SDSS-XDQSO quasar targeting catalog contains
160,904,060 point-sources with dereddened $i$-band magnitude between
17.75 and 22.45 mag in the 14,555 $deg^2$ of imaging from SDSD DR8.
For our training sets, we just select the objects (No. 301,043) in
the subarea of Stripe-82 except those predicted as quasars by
SDSS-XDQSO (the probability of quasars $> 0.5$).

The test set are composed of two parts. The first one is one tenth
(No. 95,202) of quasar Catalogue V and the second one is one thirds
(No. 160,294) of {\it SpecPhotoAll} table in SDSS DR8 which has been
cleaned in the above paragraph.

\section{METHOD}
\label{sec:method}
\subsection{SVM}
\label{subsec:svm} 
\begin{figure}
\centering
\includegraphics[width=0.4\textwidth,angle=0]{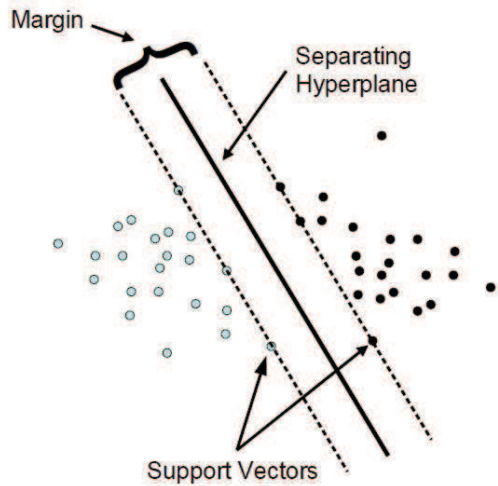}
\caption{This is a linear separable case of SVM.}
\label{fig:SV}
\end{figure}

Support Vector Machines (SVM), proposed by Vapnik
(1995), is derived from the theory of structural risk minimization
which belongs to statistical learning theory. The core idea of SVM
is to map input vectors into a high-dimensional feature space and
construct the optimal separating hyperplane in this space. SVM aims
at minimizing an upper bound of the generalization error through
maximizing the margin between the separating hyperplane and the
data. Basically, we are looking for the optimal separating
hyperplane between the two classes by maximizing the margin between
the classes' closest points. In Figure \ref{fig:SV} \footnote{This
figure is plotted by David 2001} points lying on the boundaries are
called support vectors and it means that SVM just uses the most
representative points to construct a classifier not using all of
them.

For a given training set belonging to two different classes is often called positive class and negative
class (or plus class and minus class),
\begin{equation}
T={(\vec{x}_1,y_1),\ldots,(\vec{x}_n,y_n)},\qquad
\vec{x}_i\in\mathbf{R^N},y_i\in{\{-1,+1\}}
\end{equation}
SVM learns linear threshold functions of the type.
\begin{equation}
h(\vec{x}_i) = sign\{\vec{\omega}\cdot\vec{x}_i+b\}=\left\{
\begin{aligned}
+1,& ~~~\mbox{if}~~~\vec{\omega}\cdot\vec{x}_i+b > 0\\
-1,& ~~~\mbox{else}
\end{aligned}
\right.
\end{equation}
Each linear threshold function corresponds to a hyperplane in a
feature space and the side of the hyperplane on which an example
$\vec{x}_i$ lies determines the classified result by the function
$h(\vec{x}_i)$. If the training data can be separated by at least
one hyperplane $h'$, the optimal hyperplane with maximum margin can
be found by minimizing

\begin{equation}
\label{eq:soft_SVMs}
F(\vec{\omega},\vec{\xi})=\frac{1}{2} (\vec{\omega} \cdot \vec{\omega})+C\sum_{i=1}^{n}\xi_i
\end{equation}
which subjects to
\begin{equation}
y_i[(\vec{\omega} \cdot \vec{x}_i)+b]\ge 1-\xi_i \qquad i=1,\ldots,n
\end{equation}
\begin{equation}
\xi_i > 0 \qquad i=1,\ldots,n
\end{equation}

The factor $C$ is used to trade off training error against model
complexity and $\xi$ are slack variables responding to the wrong
prediction. In practice, we would like to penalize the errors on
positive examples (quasars) stronger than errors on negative
examples (stars), because we are much more interested in quasars
than stars and the quantity of stars is often much larger than that
of quasars. Morik et~al. (1999) modified the Eq.~\ref{eq:soft_SVMs}
through minimizing
\begin{equation}
\label{eq:asymmetric_SVMs}
F(\vec{\omega},\vec{\xi})=\frac{1}{2} (\vec{\omega} \cdot \vec{\omega})+C_{-+}\sum_{i:y_i=+1}\xi_i+C_{+-}\sum_{j:y_j=-1}\xi_j
\end{equation}
which is constrained by
\begin{equation}
y_i[(\vec{\omega} \cdot \vec{x}_i)+b]\ge 1-\xi_k \qquad  k=1,\ldots,n
\end{equation}
We can use the both factors $C_{-+}$ and $C_{+-}$ to control the
cost of false positives versus false negatives and get the result
that we focus on. The books (Vapnik 1995; Vapnik 1998) contain
excellent description of SVM and the article written by Burges
(1998) provides a good tutorial on it. In this paper, we adopt
$SVM^{light}$ coded by Joachims
(2002)\footnote{http://svmlight.joachims.org/} which is an
implementation of SVM in C language with many extensional and
additional softwares, moreover this code provides various model
parameters including kernel functions for us to tune.

\subsection{Build a SVM classification system}
\label{subsec:svm_sys}

\begin{figure*}
\centering
\includegraphics[width=0.8\textwidth,angle=0]{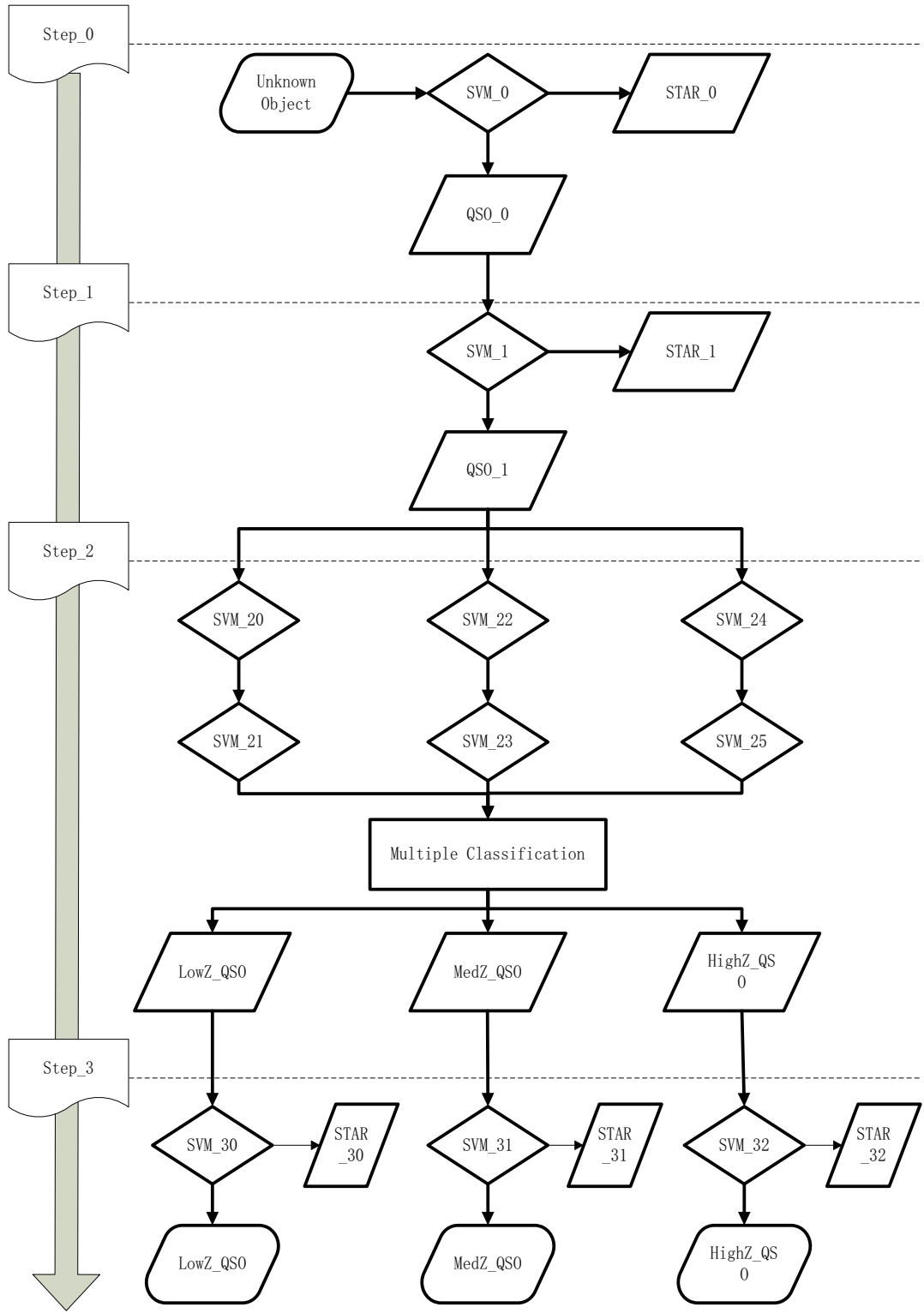}
\caption{The scheme of a SVM classification system. A total of eleven classifiers (SVM\_0, SVM\_1, SVM\_20, SVM\_21, SVM\_22, SVM\_23, SVM\_24, SVM\_25, SVM\_30, SVM\_31 and SVM\_32) for separating quasars from stars are trained by SVM. After being processed by these classifiers, any unknown sample
will be classified in one of the four categories, namely LowZ\_QSO, MedZ\_QSO, HighZ\_QSO and stars ( STAR\_0, STAR\_1, STAR\_2,
STAR\_30, STAR\_31 and STAR\_32).}
\label{fig:SVM-draft}
\end{figure*}

In this chapter, we discuss how to use several SVM models to build a
SVM classification system for selecting quasar candidates in detail.
The input pattern of SVM is a combination of photometric magnitudes
and colors, just like the combination (psfMag\_u-psfMag\_g,
psfMag\_g-psfMag\_r, psfMag\_r-psfMag\_i, psfMag\_i-psfMag\_z,
psfMag\_r) mentioned in Gao et al. (2008). All magnitudes in this
combination have been corrected by the map of Schlegel et al.
(1998). In Figure~\ref{fig:SVM-draft}, we give the scheme of the SVM
classification system with four steps and eleven models. Although
many data mining algorithms have been successfully applied on this
problem, most of them solved it only with one classifier. Actually
this is very hard for one model to include all information at the
same time and limits the performance of a classifier. Our idea is
that we divide this task into several relative simple subtasks and
conquer them respectively. The work of Step\_0 (SVM\_0) is about
eliminating the stars that are apparently different from quasars.
Step\_1 (SVM\_1) is mainly to separate quasars from the confusing
stars. These two steps are the foundation of this system and many
other authors combine the both together or just deal with one of
them. The duty of Step\_2 (from SVM\_20 to SVM\_25) is to divide the
quasar candidates into three subclasses. Finally, Step\_3 (SVM\_30,
SVM\_31 and SVM\_32) can make a further clean of the quasar
candidates of each subclass and improve the prediction accuracies of
the subclasses much higher.
%The input pattern and model parameters used to construct this system will be discussed in Chapter \ref{sec:pef}.

For constructing the classifier of SVM\_0, the above mentioned
training samples of stars and quasars in Section~\ref{sec:data} will
be used to build a classifier. The training samples of stars are
adopted from two thirds of spectroscopically confirmed stars in SDSS
DR8, photometric stars in SDSS DR8 with dereddened $i$-band
magnitude $< 17.75$, photometric stars in SDSS-XDQSO with dereddened
$i$-band magnitude between 17.75 and 22.45 and the probability-XDQSO
less than 0.5. Nine tenths of spectral identified quasars in
Schneider's Catalog V are randomly sampled and taken as the training
sample of quasars. Considering the small sample of quasars, we don't
put constraint on quasars in the scope of the subarea of Stripe-82.
Generally when the completeness is higher, the efficiency is lower.
Since our primary goal in this session is to weed out most stars
(i.e. STAR\_0) which are apparently different from quasars and easy
to be eliminated, the low efficiency can be accepted. In Table
\ref{tab:tr_data}, we list all training sets used in each
classifier. Many confusing stars will be mixed into our quasar
candidates (QSO\_0) of SVM\_0 in this step but we reserve quasars as
many as possible.

\begin{table}
\begin{center}
\caption{The number of training data used in each SVM model. Positive and negative separately denote what role played by
the corresponding quasars or stars for constructing a classifier. From SVM\_20 to SVM\_25 just use quasars as the positive and
negative samples because their functions are to classify quasar candidates into the three subcategories: LowZ\_QSO, MedZ\_QSO
and HighZ\_QSO. }\label{tab:tr_data}
\begin{tabular}{lcc}
\hline
Model & Positive (QSO) &  Negative (Star) \\
      & No.    & No.  \\
\hline
SVM\_0  & 95,202  & 442,309\\
SVM\_1  & 93,773  & 6,474\\
SVM\_30 & 79,635  & 1,381\\
SVM\_31 & 10,396  & 95\\
SVM\_32 &  3,001  & 105\\
\hline
& LowZ\_QSO & MedZ \& HighZ\_QSO\\
      & No.    & No.  \\
\hline
SVM\_20 & 79,421 & 15,781\\
\hline
& MedZ\_QSO & HighZ\_QSO\\
     & No.    & No.  \\
\hline
SVM\_21 &  12,610 & 3,171\\
\hline
& MedZ\_QSO &  LowZ \& HighZ\_QSO\\
      &  No.    &  No.  \\
\hline
SVM\_22 & 12,610 & 82,529\\
\hline
& LowZ\_QSO & HighZ\_QSO\\
&  No.    &  No.  \\
\hline
SVM\_23 &  79,421 & 3,171\\
\hline
& HighZ\_QSO & LowZ \& MedZ\_QSO\\
      &  No.    &  No.  \\
\hline
SVM\_24 & 3,171 & 92,031\\
\hline
& LowZ\_QSO & MedZ\_QSO\\
& SV No.    & SV No.  \\
\hline
SVM\_25 &  79,421 & 12,610\\
\hline
\end{tabular}
\end{center}
\end{table}

After getting the SVM\_0 model, we use it to process the data set
composed of two thirds of spectroscopically identified stars in SDSS
DR8 and nine tenths of quasars in Schneider's QSO Catalogue V. The
objects labeled as quasar candidates (QSO\_0) by SVM\_0 contain most
of genuine quasars (No. 94,603) and many confusing stars (No.
6,474). These objects will be used to form the training set for
SVM\_1. We directly discard the objects marked as STAR\_0 ( 314,110
stars and 599 quasars) by SVM\_0 because the responsibility of
SVM\_1 is to distinguish the objects that can not be solved by
SVM\_0. When SVM\_1 model is applied to QSO\_0, many confusing stars
will be removed out of it.

In order to divide the quasar candidates into three subclasses:
LowZ\_QSO (low-redshift quasars), MedZ\_QSO (medium-redshift
quasars) and HighZ\_QSO (high-redshift quasars), there is a multiple
classification with three branches needed to be built using nine
tenths of quasars in Schneider's QSO catalog V without adding any
star sample. QSO\_1 obtained by SVM\_1 will be processed through
three branches, each of them is a two-layer classifier and then the
objects in QSO\_1 will be marked as the subclass that gets the most
votes. In Figure \ref{fig:SVM-draft}, for example, there are SVM\_20
and SVM\_21 in the first branch to discriminate LowZ\_QSO, MedZ\_QSO
and HighZ\_QSO. SVM\_20 classifies LowZ\_QSO from MedZ\_QSO and
HighZ\_QSO and then SVM\_21 distinguishes MedZ\_QSO from HighZ\_QSO.
After processing by the two models, the quasar candidates will get a
subcategory and the corresponding prediction value made by SVM. In
the second branch, SVM\_22 deals with MedZ\_QSO vs. LowZ/HighZ\_QSO
and SVM\_23 handles LowZ\_QSO vs. HighZ\_QSO. In the third branch,
SVM\_24 deals with HighZ\_QSO vs. LowZ/MedZ\_QSO and SVM\_25 handles
LowZ\_QSO vs. MedZ\_QSO. When the object gets the same vote with
LowZ\_QSO, MedZ\_QSO and HighZ\_QSO, the category with the maximum
absolute SVM prediction value will be assigned to this object. The
maximum absolute SVM prediction value of one quasar candidate means
that it is farthest away from the optimal separate hyperplane and it
is more likely to belong to this class.

\begin{table}
\begin{center}
\caption{The number of support vectors (SV) used in each SVM model.
Positive denotes which type of objects marked as \textit{plus} and
negative denotes which type of objects marked as \textit{minus} in a
SVM model. }\label{tab:sv}
\begin{tabular}{lcc}
\hline
Model & Positive (QSO) &  Negative (Star) \\
      & SV No.     & SV No.  \\
\hline
SVM\_0  &  3,641 & 3,849\\
SVM\_1  &  5,480 & 5,424\\
SVM\_30 &  1,494 & 1,381\\
SVM\_31 &    168 &   95\\
SVM\_32 &    167 &   105\\
\hline
& LowZ\_QSO & MedZ \& HighZ\_QSO\\
      & SV No.    & SV No.  \\
\hline
SVM\_20 & 4,835 & 4,889\\
\hline
& MedZ\_QSO & HighZ\_QSO\\
& SV No.    & SV No.  \\
\hline
SVM\_21 &  666 & 679\\
\hline
& MedZ\_QSO &  LowZ \& HighZ\_QSO\\
      & SV No.    & SV No.  \\
\hline
SVM\_22 & 4,145 & 4,195\\
\hline
& LowZ\_QSO & HighZ\_QSO\\
& SV No.    & SV No.  \\
\hline
SVM\_23 &  665 & 662\\
\hline
& HighZ\_QSO & LowZ \& MedZ\_QSO\\
      & SV No.    & SV No.  \\
\hline
SVM\_24 & 442 & 439\\
\hline
& LowZ\_QSO & MedZ\_QSO\\
& SV No.    & SV No.  \\
\hline
SVM\_25 &  4,835 & 4,828\\
\hline

\end{tabular}
\end{center}
\end{table}

The remaining three SVM models are SVM\_30, SVM\_31 and SVM\_32.
Their functionalities are to eliminate some very indistinguishable
stars from LowZ\_QSO, MedZ\_QSO and HighZ\_QSO, respectively. The
main idea is that when quasar candidates selected out by SVM\_0 and
SVM\_1, the classifiers utilize the general characteristics of
quasars and stars. In the subcategory, we can make use of its own
characteristics to get a more pure quasar set. Therefore, the
training sets for the three models are based on the positive class
(QSO\_1) extracted from the data set composed of two thirds of
spectroscopically identified stars in SDSS DR8 and nine tenths of
quasars in Schneider's QSO Catalogue V processed by SVM\_0 and
SVM\_1 and then this positive class will be divided into three
segments by multiple classification. The three segments will be used
for generating SVM\_30, SVM\_31 and SVM\_32 separately and each of
them includes some indistinguishable stars that can not be simply
weeded out by SVM\_0 and SVM\_1. Through these three models, a small
amount of star contaminants is removed. It is also noticed that the
risk of misclassifying a number of genuine quasars into star
contaminants exists especially for high-redshift quasars.

In Table \ref{tab:sv}, we list the number of support vectors used in
each of the SVM models. Obviously SVM does not need to use all
samples to construct a classifier because it only uses the samples
located on the optimal separating hyperplane in a high-dimensional
feature space. The number of support vectors reflects the complexity
of the problem solved by a classifier. Although the training set of
SVM\_0 is the largest one, the number of support vectors is small
because most of stars can be easily separated from quasars. The most
hard work belongs to SVM\_1 and this model includes 5,480 quasars
and 5,424 stars as support vectors because many quasars and stars
are very similar even in a high-dimensional feature space.

\section{THE PERFORMANCE OF SVM}
\label{sec:pef}

\begin{table*}
\begin{center}
\caption{The evaluation of the ability of SVM for predicting the quasar subclasses: LowZ\_QSO, MedZ\_QSO and HighZ\_QSO using the test quasar samples retained in QSO\_1.
The number 8543 shows how many true low-redshift quasars are correctly predicted in LowZ\_QSO. The value 97.25\% represents the efficiency (E.)
of predicted LowZ\_QSO to true low-redshift quasars. The value 98.66\% denotes the completeness (C.) of true low-redshift quasars in all predicted
quasar sample. }\label{tab:redshift}
\begin{tabular}{llllllllll}
\hline
Predicted Class & \multicolumn{3}{c}{True LowZ\_QSO} & \multicolumn{3}{c}{True MedZ\_QSO} & \multicolumn{3}{c}{True HighZ\_QSO} \\
& No. & E.(\%) & C.(\%) & No. & E.(\%) & C.(\%) & No. & E.(\%) & C.(\%) \\
\hline
LowZ\_QSO  & 8543 & 97.25 & 98.66 & 242  & 2.75  & 18.11 & 0  & 0.00  & 0.00  \\
MedZ\_QSO  & 113  &  9.37 & 1.31  & 1083 & 89.80 & 81.06 & 10  & 0.83  & 3.12  \\
HighZ\_QSO & 3    & 0.93  & 0.03  & 11   & 3.40  & 0.82  & 310 & 95.68 & 96.88 \\
\hline
\end{tabular}
\end{center}
\end{table*}

\begin{table*}
\begin{center}
\caption{The number and the fraction of stars mixed into our
predicted categories LowZ\_QSO, MedZ\_QSO and HighZ\_QSO,
respectively. The column test (No.) shows the number of stars that
would be used to test in our method. The column Step\_0 (No.) means
how many stars can not be removed by Step\_0 of SVM. For A0 stars,
the number 52 of A0 in LowZ\_QSO indicates how many A0 stars finally
can not be eliminated by Step\_3. The decimal number 6.92 (i.e.
52/751) represents the contaminant percentage of A0 stars occupied
in the whole contaminant sample set (751 stars can not be correctly
classified by our method). The another decimal number 0.29 (i.e.
52/17953) means that this percentage of A0 stars in the whole A0
stars will be misclassified as LowZ\_QSO. None of Carbon\_lines
stars mixed into any one of the three classes and we use "--.--" to
represent the percentage is null. }\label{tab:STAR}
\begin{tabular}{lcccrrcrrcrr}
\hline
Star Subclass&Test  & Step\_0 & \multicolumn{3}{c}{LowZ\_QSO} & \multicolumn{3}{c}{MedZ\_QSO} & \multicolumn{3}{c}{HighZ\_QSO} \\
&No. & No. & Step\_3 No. & P.(\%) & P.(\%) & Step\_3 No. & P.(\%) & P.(\%) & Step\_3 No. & P.(\%) & P.(\%) \\
\hline
A0 &17953 &349 & 52  & 6.92 & 0.29  & 15& 2.00 & 0.08 & 0 & 0.00 & 0.00 \\
A0p &348 &6    & 3   & 0.40 & 0.86  & 0 & 0.00 & 0.00  & 0 & 0.00 & 0.00 \\
B6 &138  &20   & 9   & 1.20 & 6.52  & 2 & 0.27 & 1.45 & 0 & 0.00 & 0.00 \\
B9 &200  &13   & 4   & 0.53 & 2.00  & 1 & 0.13 & 0.50 & 0 & 0.00 & 0.00 \\
CV &594  &368  & 215 & 28.63& 36.20 & 2 & 0.27 & 0.34 & 0 & 0.00 & 0.00 \\
Carbon &79 &5  & 1   & 0.13 & 1.27  & 0 & 0.00 & 0.00 & 0 & 0.00 & 0.00 \\
CarbonWD &36 &30& 17 & 2.26 & 47.22 & 0 & 0.00 & 0.00 & 0 & 0.00 & 0.00 \\
Carbon\_lines &195 &36 & 0 & --.-- & --.-- & 0 & --.-- & --.-- & 0 & --.-- & --.-- \\
F2 &4170 &19   & 0  & 0.00 & 0.00  & 1 & 0.13  & 0.02 & 0 & 0.00 & 0.00 \\
F5 &27888 &180 & 10 & 1.33 & 0.04  & 11& 1.46  & 0.04 & 0 & 0.00 & 0.00 \\
F9 &34262 &133 & 0  & 0.00 & 0.00  & 5 & 0.67  & 0.01 & 1 & 0.13 & 0.00 \\
G0 &3289  &13  & 0 & 0.00  & 0.00  & 2 & 0.27  & 0.06 & 0 & 0.00 & 0.00 \\
G2 &8399  &39  & 1 & 0.13  & 0.01  & 6 & 0.80  & 0071 & 0 & 0.00 & 0.00 \\
G5 &1     &0   & 0 & --.-- & --.-- & 0 & --.-- & --.-- & 0 & --.-- & --.-- \\
K1 &8505  &141 & 1 & 0.13  & 0.01  & 1 & 0.13  & 0.01  & 1 & 0.13 & 0.01 \\
K3 &8997  &365 & 3 & 0.40  & 0.03  & 2 & 0.27  & 0.02  & 4 & 0.53 & 0.04 \\
K5 &7957  &241 & 1 & 0.13  & 0.01  & 0 & 0.00  & 0.00  & 14& 1.86 & 0.18 \\
K7 &5430  &99  & 2 & 0.27  & 0.04  & 0 & 0.00  & 0.00  & 11& 1.46 & 0.20 \\
L0 &18    &0   & 0 & --.-- & --.-- & 0 & --.-- & --.--& 0 & --.-- & --.-- \\
L1 &14    &0   & 0 & --.-- & --.-- & 0 & --.-- & --.--& 0 & --.-- & --.-- \\
L2 &41    &0   & 0 & --.-- & --.-- & 0 & --.-- & --.--& 0 & --.-- & --.-- \\
L3 &6     &0   & 0 & --.-- & --.-- & 0 & --.-- & --.--& 0 & --.-- & --.-- \\
L4 &9     &0   & 0 & --.-- & --.-- & 0 & --.-- & --.--& 0 & --.-- & --.-- \\
L5 &10    &0   & 0 & --.-- & --.-- & 0 & --.-- & --.--& 0 & --.-- & --.-- \\
L5.5 &56  &9   & 4 & 0.53  & 7.14  & 1 & 0.13  & 1.79 & 0 & 0.00 & 0.00 \\
L9 &66    &9   & 7 & 0.93  & 10.61 & 0 & 0.00 & 0.00  & 0 & 0.00 & 0.00 \\
M0 &3665  &53  & 0 & 0.00  & 0.00  & 0 & 0.00 & 0.00  & 12& 1.60 & 0.33 \\
M0V &604  &8   & 0 & --.-- & --.--& 0 & --.-- & --.-- & 0 & --.--& --.-- \\
M1 &3442  &30  & 1 & 0.13  & 0.03 & 0 & 0.00 & 0.00   & 3 & 0.40 & 0.09 \\
M2 &4922  &22  & 2 & 0.27  & 0.04 & 0 & 0.00 & 0.00   & 5 & 0.67 & 0.10 \\
M2V &162  &1   & 0 & --.-- & --.--& 0 & --.-- & --.-- & 0 & --.--& --.-- \\
M3 &4604  &34  & 1 & 0.13  & 0.02 & 0 & 0.00 & 0.00   & 2 & 0.27 & 0.04 \\
M4 &3099  &29  & 0 & 0.00  & 0.00 & 0 & 0.00 & 0.00   & 1 & 0.13 & 0.03 \\
M5 &1947  &14  & 0 & --.-- & --.--& 0 & --.-- & --.-- & 0 & --.-- & --.-- \\
M6 &2669  &8   & 0 & --.-- & --.--& 0 & --.-- & --.-- & 0 & --.-- & --.-- \\
M7 &1011  &2   & 2 & 0.27  & 0.20 & 0 & 0.00 & 0.00   & 0 & 0.00 & 0.00 \\
M8 &494   &0   & 0 & --.-- & --.--& 0 & --.-- & --.-- & 0 & --.-- & --.-- \\
M9 &360   &0   & 0 & --.-- & --.--& 0 & --.-- & --.-- & 0 & --.-- & --.-- \\
O &107    &5   & 2 & 0.27  & 1.87 & 0 & 0.00 & 0.00 & 0 & 0.00 & 0.00 \\
OB &342   &6   & 3 & 0.40  & 0.88 & 0 & 0.00 & 0.00 & 0 & 0.00 & 0.00 \\
T2 &100   &17  & 3 & 0.40  & 3.00 & 0 & 0.00 & 0.00 & 0 & 0.00 & 0.00 \\
WD &4070  &1023 & 295 & 39.28 & 7.25 & 6 & 0.80 & 0.15 & 0 & 0.00 & 0.00 \\
WDmagnetic &35&11& 3  & 0.40  & 8.57 & 0 & 0.00 & 0.00 & 0 & 0.00 & 0.00 \\
Total &160294 &3338 & 642 & 85.49 & 0.40 & 55 & 7.31 & 0.03 & 54 & 7.20 & 0.03  \\
\hline
\end{tabular}
\end{center}
\end{table*}

\begin{table}
\begin{center}
\caption{The final performances of SVM classification system. The efficiency 90.62\% of LowZ\_QSO represents the proportion of low-redshift quasars vs. other quasars and stars. The completeness 97.35\% of it shows how many low-redshift quasars will be recovered from all genuine low-redshift quasars.) }\label{tab:finalpef}
\begin{tabular}{lccrr}
\hline
Predicted Class & Efficiency & Completeness \\
\hline
LowZ\_QSO       & 90.62\% & 97.35\% \\
MedZ\_QSO       & 85.88\% & 74.35\% \\
HighZ\_QSO      & 82.01\% & 89.08\% \\
\hline
Total           & 93.21\% & 97.49\% \\
\hline
\end{tabular}
\end{center}
\end{table}

The performance of this SVM classification system is tested by the
test set including one third of spectroscopically confirmed stars
(No. 160,294 ) in SDSS DR8 and one tenth of quasars (No. 10,581) in
Schneider's QSO Catalogue V. This classification system has been
described in Chapter \ref{subsec:svm_sys}. The model parameters for
each classifier can be found in Appendix A.

After this test set gets through Step\_0, 10,399 quasars and 3,338
stars are kept in QSO\_0. The efficiency of SVM reaches 75.70\% and
the completeness of it is 98.28\%. By means of this model SVM\_0,
most of stars (No. 156,956) are weed out and a small amount of
quasars (No. 182) are just lost. These weeded stars are so obviously
discriminated from quasars that they are easy to remove. It is
concluded from the large number (No. 156,956) that such stars occupy
the majority of stars. Therefore this step is necessary and helpful
to clear away the pollution of most of stars. Usually in previous
literatures, this step is lack. They focused on separating confusing
stars from quasars. This is the reason that the number of their
targeting quasars is rather large.

%There is a little strange for the high efficiency produced by SVM in this step for some readers. The reason is that the test samples of quasar is relatively large. Fortunately, we can infer the possible genuine efficiency by reducing the size of quasar samples to keep the ratio of QSO vs. STAR as 1:100. Consequently, the new efficiency of SVM in this step is 58.16\% and the completeness of it is also 98.29\%. In the following experiments, we will just use this test set to assess the performance of our system.
% low 61 med 101 high 19
% 0.69\% 6.95\%  5.46\%
% low 33 med 28 high 30
% 1.05\% 8.88\% 14.08\%

In Step\_1 (SVM\_1), it will eliminate the confusing stars from
QSO\_0 and almost two thirds of stars (No. 2,583) are selected out
with 85 quasars lost. The efficiency and the completeness of SVM\_1
becomes 93.18\% and 97.49\%, respectively. Apparently, this step can
further contribute to avoid the pollution of many confusing stars,
meanwhile, a small number of quasars are inevitably missing. Perhaps
adding infrared information from UKIDSS database (Lawrence et~al.
2007) into the SVM model or directly using some color-color criteria
(e.g. Wu et~al. 2010, 2011) are helpful to recover some missing
medium and high-redshift quasars in this step.

When computing the performances of SVM to classify low, medium and high-redshift quasars, stars are not considered in Step\_2.
In Table \ref{tab:redshift}, the efficiency of these three subclasses is 97.25\%, 89.80\% and 95.68\%, separately, and the completeness of them is 98.66\%,
81.06\% and 96.88\%, respectively. The matrix of Table \ref{tab:redshift} proves that SVM can obtain good performance with multiple classification and 18.11\% of medium-redshift quasars are easily classified into the low-redshift quasars. Perhaps given data from more bands, discrimination of LowZ\_QSO and MedZ\_QSO becomes more efficient.

\begin{figure}
\centering
\includegraphics[width=0.5\textwidth,angle=0]{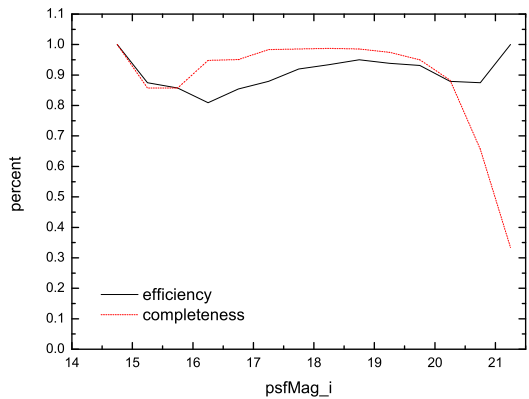}
\caption{Efficiency and completeness as a function of magnitude $i$,
solid line for efficiency, dotted line for completeness.}
\label{fig:perf}
\end{figure}

Until SDSS DR8 release, SDSS begins to provide a detailed subclasses
of stars. The number of subclasses amounts to 43 considering each
spectroscopically confirmed star. Table \ref{tab:STAR} shows that
the number and the fraction of the 43 subclasses of stars are mixed
into our predicted categories LowZ\_QSO, MedZ\_QSO and HighZ\_QSO,
respectively and provides what type of stars may mostly be mixed
into quasars by SVM after Step\_3. It is found that WD (45.95\%), CV
(33.49\%), A0 (8.10\%), CarbonWD (2.65\%) and F5 (1.56\%) can easily
be misclassified as LowZ\_QSO. Most of contaminants in MedZ\_QSO are
A0(27.27\%), F5 (20.00\%), G2 (10.91\%), WD (10.91\%) and F9
(9.09\%). A0 and F5 stars can be easily misclassified into both
low-redshift and medium-redshift quasars. The situation of
HighZ\_QSO is different that contaminants mainly come from K or M
stars. The number in the parenthesis of Table \ref{tab:STAR}
represents the misclassified stars before Step\_3. We can find some
information about Step\_3 that SVM\_30 can weed out some A0, CV and
WD stars, SVM\_31 mainly eliminate some A0 and F5 stars. Finally,
the efficiency and the completeness of the SVM classification system
is 93.21\% and 97.49\%, respectively. In Table 4, the final
efficiency of these three subclasses is 90.62\% (LowZ\_QSO vs. other
quasars and stars), 85.88\% (MedZ\_QSO vs. other quasars and stars)
and 82.01\% (HighZ\_QSO vs. other quasars and stars) separately and
the completeness of them is 97.35\% (correctly predicted LowZ\_QSO
vs. all genuine LowZ\_QSO), 74.35\% (correctly predicted MedZ\_QSO
vs. all genuine MedZ\_QSO) and 89.08\% (correctly predicted
HighZ\_QSO vs. all genuine HighZ\_QSO). For Carbon\_lines, G5, L0,
L1, L2, L3, L4, M0V, M2V, M5, M8, and M9 stars, none of them is
misclassified into quasars. Figure \ref{fig:perf} shows the
efficiency and completeness as a function of magnitude $i$. However,
the trend with magnitude $i < 16.5$  is unreliable for the number of
sample is just a few during this magnitude range. The real trend
needs a larger sample to deduce. As magnitude $i > 16.5$, the number
of sample increases to hundreds or more than hundreds. Therefore the
tendency in this range is credible. No matter for efficiency or
completeness, the run is steady during the range $17 < i < 19.5$,
then goes down beyond $i$=19.5. That the efficiency goes up and
completeness declines beyond $i$=20.2 is unreliable due to small
sample in this magnitude range and magnitude limit.

\section{QUASAR CANDIDATE SELECTION}

\label{sec:qso_sel} Through the above experiments, the SVM
classification system proved applicable and reasonable to select
quasar candidates from large sky survey projects. In order to
further demonstrate the efficiency of this system, the comparison
with the work of Bovy et~al. (2011) has been done as follows.
XD-sources is an unknown point-sources produced by Bovy et~al.
(2011) and we use it to generate a part of the quasar input catalog
for Guoshoujing Telescope (LAMOST) with our SVM system in the pilot
survey. SDSS-XDQSO quasar targeting catalog can be directly
downloaded from the web page
\footnote{http://data.sdss3.org/sas/dr8/groups/boss/photoObj/xdqso/xdcore}
provided by Bovy et~al. (2011). It includes 160,904,060
point-sources with dereddened $i$-band magnitude between 17.75 and
22.45 mag from SDSS DR8. The flag cuts for every source in this
catalog have been used to filter unqualified ones. The detailed
information about these flag cuts can be found in the Appendix A of
the paper of Bovy et~al. (2011). XDQSO technique has been applied on
all objects in this catalog to provide the types and probabilities
of them. Objects which satisfy the XDQSO probability cut
P(XDQSO\_MedZ)$>$0.424 will be selected as CORE targets in SDSS-III
BOSS.

The Guoshoujing Telescope
(LAMOST)\footnote{http://www.lamost.org/website/en} is an innovative
reflecting Schmidt telescope with 4 meter effective mirror size, 20
square degree field of view and 4000 fibres. It will perform most
efficient optical spectroscopic sky survey. It entered the pilot
survey phase in the end of 2011 and will carry out the regular
survey in this year. Careful preparation of the input catalog for
LAMOST is important for the scientific output of LAMOST. Since
LAMOST has no own photometric data, the photometric data from other
survey projects should be depended on, such as SDSS, UKIDSS, WISE,
GALEX. In the pilot survey, two chunks are selected (-$45^\circ <
\alpha_{\mathrm{J2000}}<+60^\circ$ and -$1^\circ.5 <
\delta_{\mathrm{J2000}} < +8^\circ.5$ ; $+180^\circ <
\alpha_{\mathrm{J2000}}<+210^\circ$ and $+12^\circ <
\delta_{\mathrm{J2000}} < +23^\circ$). We use the SVM classification
system to select quasar candidates and compare our result with the
targets selected by XDQSO technique in the both chunks.

Our SVM classification system obtains 64,660 targets in chunk1 and 29,520 targets in chunk2. Table \ref{tab:com} indicates that the selected
quasar candidates by SVM overlap those by XDQSO in different probability ranges. Most of targets selected by SVM are covered by XDQSO especially for the highest probability
($P>0.99$) of XDQSO. In chunk1 and chunk2, 57.38\% and 87.01\% quasar candidates selected by SVM are also targeted by XDQSO. This can
make the targets selected by SVM to a higher confidence. Table \ref{tab:cof} indicates that the consistency of the two methods for
classifying targets into three subcategories: low-redshift quasars, medium-redshift quasars and high-redshift quasars, and that the
difference between the two methods is small except that some targets predicted as LowZ\_QSO by SVM are classified by XDQSO
as MedZ\_QSO.

\begin{figure*}
\centering
\includegraphics[width=1.0\textwidth,angle=0]{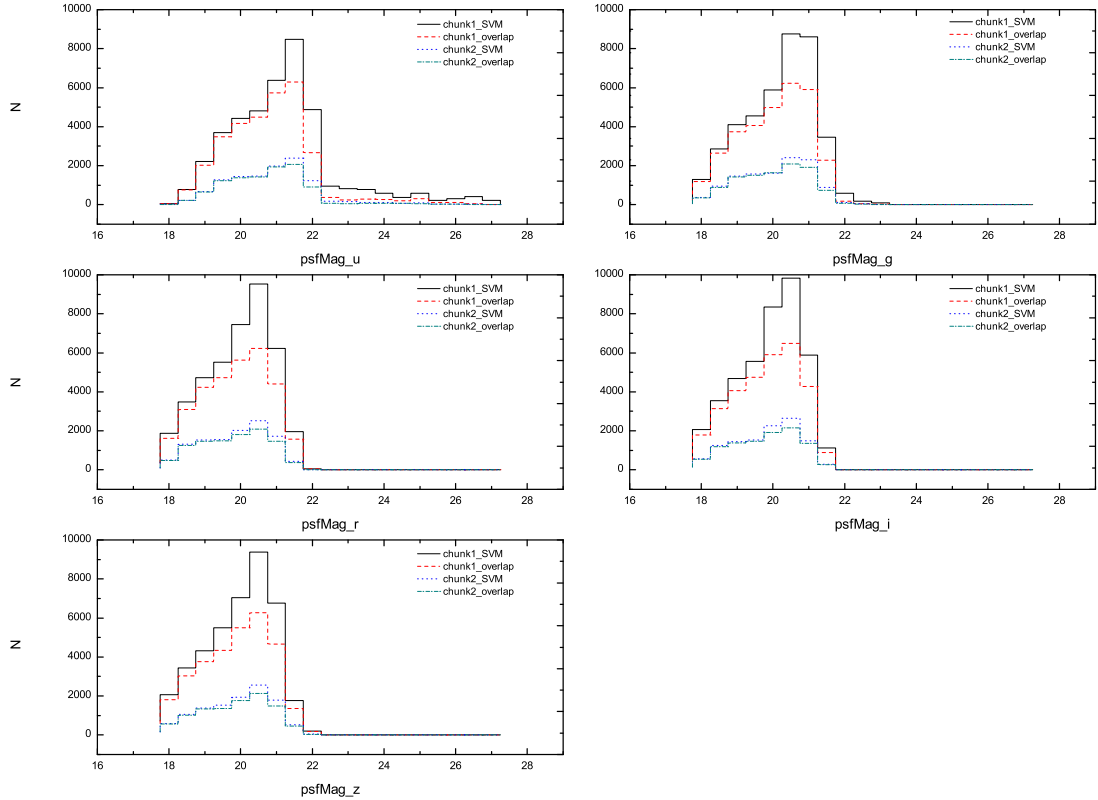}%compare-result.eps}
\caption{Comparison of the quasar candidates predicted only by SVM
with those predicted both by SVM and XDQSO. The predictions of the
two chunks are plotted separately. Chunk1\_overlap and
chunk2\_overlap represent the quasar candidates selected both by SVM
and XDQSO in these two pilot areas respectively. Solid line
represents chunk1\_SVM, long dashed line for chunk1\_overlap, dotted
line for chunk2\_SVM, long dashed dotted line for chunk2\_overlap.}
\label{fig:com-res}
\end{figure*}

Actually, the amount of targets selected by this system is smaller
than that by XDQSO because we want to get the higher predicted
efficiency of quasars. In Figure \ref{fig:com-res}, the predicted
results of SVM in the two chunks as well as the overlaps of SVM and
XDQSO are shown. It is found that the prediction of SVM coincides
with that of XDQSO , especially in psfMag\_i$<$20.0. The main reason
of the difference of SVM and XDQSO in psfMag\_i$>$20.0 maybe come
from that the training sample includes so small a number of faint
celestial objects that the ability of this system to recognize these
objects is weak.

\begin{table}
\begin{center}
\caption{The number per $deg^2$ of quasar candidates selected by SVM
overlaps those selected by XDQSO in chunk1 and chunk2. The number in
the parenthesis is produced by XDQSO.  }\label{tab:com}
\begin{tabular}{lrr}
\hline
XDQSO Probability  & Chunk1 $deg^2$ & Chunk2 $deg^2$\\
\hline
$0.990\le P     $ & 12.7 (15.9) & 31.0 (40.0)\\
$0.950\le P<0.990$ & 9.4 (12.6) & 22.7 (31.3)\\
$0.900\le P<0.950$ & 4.0 (7.3) & 8.0  (16.9)\\
$0.850\le P<0.900$ & 2.1 (5.2)& 4.1  (12.1)\\
$0.800\le P<0.850$ & 1.4 (4.7)& 2.8  (10.9)\\
$0.750\le P<0.800$ & 1.2 (4.4)  & 2.0  (10.6)\\
$0.700\le P<0.750$ & 1.1 (4.5) & 1.7  (10.6)\\
$0.650\le P<0.700$ & 0.9 (4.7) & 1.4  (10.6)\\
$0.600\le P<0.650$ & 0.9 (5.0)  & 1.0  (10.6)\\
$0.550\le P<0.600$ & 0.8 (5.3)  & 1.1  (11.6)\\
$0.500\le P<0.550$ & 0.8 (5.9)  & 0.9  (11.7)\\
%$\qquad\quad\,\,\, P<0.500$ & 5.1 & 12.2 \\
\hline
\end{tabular}
\end{center}
\end{table}

\begin{table}
\begin{center}
\caption{The matrices of quasar candidates predicted as LowZ\_QSO, MedZ\_QSO and HighZ\_QSO by SVM and XDQSO in chunk1 (No. 3,600,423 sources ) and chunk2 (No. 1,531,240 sources). Both the matrices
reflect the agreement of the prediction of SVM and XDQSO. }\label{tab:cof}
\begin{tabular}{lccc}
\hline
Chunk1 & SVM\_LowZ & SVM\_MedZ & SVM\_HighZ\\
\hline
XDQSO\_LowZ  & 30512 & 234  &  29 \\
XDQSO\_MedZ  & 997   & 4245 &  41 \\
XDQSO\_HighZ & 8     & 14    & 1022 \\
\hline
Chunk2 & SVM\_LowZ & SVM\_MedZ & SVM\_HighZ\\
\hline
XDQSO\_LowZ  & 21022 & 122  &  6 \\
XDQSO\_MedZ  &716  & 3004 &  13 \\
XDQSO\_HighZ & 0    & 5   & 381 \\
\hline

\end{tabular}
\end{center}
\end{table}

\section{CONCLUSIONS}

\label{sec:conclusion} We have put forward a classification system
by using a hierarchy of several SVM classifiers. The above
experimental results demonstrate that single SVM classifier can not
well solve the problem of separating quasars from stars, however the
combination of some SVM classifiers gets a rather good performance.
This method can help us to select a quasar candidate set with a
relative high efficiency (93.21\%), though some actual quasars
(2.51\%) are missing in the whole process. The point we want to get
across is that the performance of this system is based on the test
sample and not on real data. In order to check the performance of
this method applied on the unknown objects, the result produced by
the method has been compared with that of the XDQSO technique. The
comparison shows that most of quasar candidates selected by the SVM
system are also recovered by XDQSO especially in the deredened
i-band magnitude $<$ 20.0. In Table~\ref{tab:cof}, actually the
prediction of SVM for subclasses of quasars also agrees with that of
XDQSO. This means that our method is an effective and feasible
approach to construct the input catalog of quasars for large
spectroscopic sky survey projects (e.g. LAMOST, SDSS ).

In the future, we plan to adopt the similar method to the XDQSO
technique to exploit whether the magnitude errors influence the
performance of the system, add the number of faint objects in the
training sample increases to improve the performance of the system
for the data set of faint objects (deredened i-band magnitude $>$
20.0). In the process of SVM\_1 where many actual quasars are
missing, we can consider some other methods to make the completeness
of quasars much higher. Each technique for quasar candidate selection has its strongness and weakness. It is difficult to say which one is better.
In terms of good efficiency, the
cross-result from different techniques to select quasar candidates
is better chosen. However, given the completeness of quasar
candidates, the combination of results from various techniques has
better be employed. We will give a much more powerful method based
on SVM to select quasar candidates for LAMOST or other projects in
the world.

\section*{Acknowledgments}
we are very grateful to the anonymous referee's constructive and
insightful comments to strengthen our paper. This paper is funded by
National Natural Science Foundation of China under grant
No.10778724, 11178021 and No.11033001, the Natural Science
Foundation of Education Department of Hebei Province under grant No.
ZD2010127 and by the Young Researcher Grant of National Astronomical
Observatories, Chinese Academy of Sciences. We acknowledgment SDSS
database. The SDSS is managed by the Astrophysical Research
Consortium for the Participating Institutions. The Participating
Institutions are the American Museum of Natural History,
Astrophysical Institute Potsdam, University of Basel, University of
Cambridge, Case Western Reserve University, University of Chicago,
Drexel University, Fermilab, the Institute for Advanced Study, the
Japan Participation Group, Johns Hopkins University, the Joint
Institute for Nuclear Astrophysics, the Kavli Institute for Particle
Astrophysics and Cosmology, the Korean Scientist Group, the Chinese
Academy of Sciences (LAMOST), Los Alamos National Laboratory, the
Max-Planck-Institute for Astronomy (MPIA), the Max-Planck-Institute
for Astrophysics (MPA), New Mexico State University, Ohio State
University, University of Pittsburgh, University of Portsmouth,
Princeton University, the United States Naval Observatory, and the
University of Washington.

\appendix

\section[]{THE MODEL PARAMETERS OF SVM }
Model parameters of SVM can greatly affect the performance of SVM for selecting quasar candidates. We generate SVM models using the
following model parameters.
\begin{eqnarray}
\begin{array}{rrllll}
{\rm A}) \;\, &  SVM\_0  & t=2 & c=100   & j=1 & g=1 \\
{\rm B}) \;\, &  SVM\_1  & t=2 & c=0.07     & j=1 & g=1 \\
{\rm C}) \;\, &  SVM\_20 & t=2 & c=0.2   & j=1 & g=1 \\
{\rm D}) \;\, &  SVM\_21 & t=2 & c=0.25 & j=1 & g=1 \\
{\rm E}) \;\, &  SVM\_22 & t=2 & c=6    & j=1 & g=1 \\
{\rm F}) \;\, &  SVM\_23 & t=2 & c=0.04 & j=1 & g=1 \\
{\rm F}) \;\, &  SVM\_24 & t=2 & c=29   & j=1 & g=1 \\
{\rm F}) \;\, &  SVM\_25 & t=2 & c=0.16 & j=1 & g=1 \\
{\rm F}) \;\, &  SVM\_30 & t=2 & c=0.12   & j=1 & g=1 \\
{\rm F}) \;\, &  SVM\_31 & t=2 & c=0.06   & j=1 & g=1 \\
{\rm G}) \;\, &  SVM\_32 & t=2 & c=0.5   & j=1 & g=1
\end{array}
\label{arr:model}
\end{eqnarray}
The parameter $t=2$ represents that SVM uses radial basis function
(RBF) kernel for deriving models. The parameter $c$ controls the
trade-off between training error and margin. The parameter $j$ in a
SVM model dominates the misclassification cost of quasars or stars.
The parameter $g$ means $\gamma$ in RBF kernel. In this work, we
just use the default value of $g$ which is equal to 1. The more
detailed information about how these parameters affect the
performance of SVM can be found in Joachims (2002). In order to
search the optimal combination of parameters $c$ and $j$, we usually
test each pair of parameters appeared in the specified sequence
which is determined by experience. For example, the first model
(SVM\_0) of this system, the parameters $c$ and $j$ are from the
values [0.2, 0.5, 1, 2, 5, 10, 20, 50, 100]. The above mentioned
parameters for each SVM model are produced by an empirical approach
because the computing time to search the optimal parameters $c$ and
$j$ is expensive. At the beginning, we just set the parameters $j$
and $g$ with default value 1. The value of parameter $c$ can be
calculated by using the sample size of stars divided by that of
quasars. This empirical method can help us to quickly get a better
parameter combination.


\begin{thebibliography}{99}
\bibitem[Abazajian et~al. 2009]{b1} Abazajian K.N. et~al., 2009,
\apjs, 182, 543
\bibitem[Abraham et~al. 2010]{b2} Abraham S., Sajeeth Philip N., Kembhavi A., Wadadekar Y.~G., Sinha R., 2010,
eprint arXiv:1011.2173
\bibitem[Aihara H. et~al. 2011]{b3} Aihara H. et~al., 2011,
\apjs, 193, 29
\bibitem[Bailer-Jones C.A.L. et~al. 2008]{b4} Bailer-Jones C.A.L., Smith K.W., Tiede C., Sordo R., Vallenari A., 2008, \mnras, 391, 1838
\bibitem[Ball N.~M. \& Brunner R.~J. 2010]{b5} Ball N.~M. \& Brunner R.~J. 2010,
IJMPD, 19, 1049
\bibitem[Borne 2009]{b6} Borne K., 2002,
eprint arXiv:0911.0505
\bibitem[Bovy et~al. 2011]{b7} Bovy J. et~al., 2011,
\apj, 729, 141
\bibitem[Burges, C.J.C. 1998]{b8} Burges, C.J.C., 1998,
PR, 167, 161
\bibitem[Carballo et~al. 2008]{b9} Carballo R., Gonz{\'a}lez-Serrano J.~I., Benn C.~R., Jim{\'e}nez-Luj{\'a}n F., 2008, \mnras, 391, 369
\bibitem[D'Abrusco R., Longo G., Walton N.A. 2009]{b10} D'Abrusco R., Longo G., Walton N.A., 2009, \mnras, 396, 223
\bibitem[Meyer D. 2001]{b101} Meyer D., 2001, R News, 23, 1(3)
\bibitem[Eisenstein D.J. et~al. 2011]{b11} Eisenstein D.J. et~al., 2011,
eprint arXiv:1101.1529
\bibitem[Gao D., Zhang Y.-X., Zhao Y.-H. 2008]{b12} Gao D., Zhang Y.-X., Zhao Y.-H., 2008,
\mnras, 386, 1417
\bibitem[Joachims T. 2002]{b13} Joachims T. 2002,
Learning to Classify Text Using Support Vector Machines, Kluwer Academic Publishers, MA
\bibitem[Kaiser \& Aussel 2002]{b14} Kaiser N. \& Aussel H., 2002,
\spie, 4836, 154
\bibitem[Kim D.-W. et~al. 2011]{b15} Kim D.-W., Protopapas P., Alcock C., Byun Y.-I., Khardon R., 2011,
ASPC, 442, 447
\bibitem[Kirtpatrik J.A. et~al. 2011]{b16} Kirtpatrik J.A. et~al., 2011,
eprint arXiv:1104.4995
\bibitem[Lawrence A. et~al. 2007]{b17} Lawrence A. et~al., 2007,
\mnras, 379, 1599
\bibitem[McPherson et~al. 2006]{b18} McPherson A.M. et~al., 2006,
\spie, 6267
\bibitem[Morik K., Brockhausen P., Joachims T. 1999]{b19} Morik K., Brockhausen P., Joachims T., 1999,
ML, CONF 16, 268
\bibitem[Richards G.T. et~al. 2002]{b201} Richards G.T. et~al. 2002,
\aj, 123, 2945
\bibitem[Richards G.T. et~al. 2004]{b20} Richards G.T. et~al. 2004,
\apjs, 155, 257
\bibitem[Richards G.T. et~al. 2009]{b21} Richards G.T. et~al. 2009,
\apjs, 180, 67
\bibitem[Ross N.P. et~al. 2011]{b22} Ross N.P. et~al. 2011,
eprint arXiv:1105.0606
\bibitem[Schlegel D.J. et~al. 1998]{b23} Schlegel D.J., Finkbeiner D.~P., Davis M., 1998,
\apj, 500, 525
\bibitem[Schlegel D.J. et~al. 2007]{b24} Schlegel D.J. et~al. 2007,
BAAS, 38, 996
\bibitem[Schneider D.P. et~al. 2010]{b22} Schneider D.P. et~al. 2010,
\aj, 139, 2360
\bibitem[Tyson 2002]{b25} Tyson J.A., 2002,
\spie, 4836, 10
\bibitem[Vapnik V.N. 1995]{b26} Vapnik V.N. 1995,
The nature of statistical learning theory, Springer US, NY
\bibitem[Vapnik V.N. 1998]{b27} Vapnik V.N. 1998,
Statistical Learning Theory, John Wiley and Sons, NY
\bibitem[Wu X.-B. \& Jia Z.-D. 2010]{b28} Wu X.-B. \& Jia Z.-D., 2010,
\mnras, 406, 1583
\bibitem[Wu X.-B. et.al. 2011]{b29} Wu X.-B., Wang R., Schmidt K.~B., Bian F., Jiang L., Fan X., 2011,
\mnras, 142, 78
\bibitem[Y\'eche C. et~al. 2010]{b30} Y\'eche C. et~al. 2010,
A\&A, 523, A14
\bibitem[York, et~al,]{b31} York, D. G., et~al., 2000, AJ, 120, 1579
\bibitem[Zhang Y.-X. \& Zhao Y.-H. 2003]{b32} Zhang Y.-X., \& Zhao Y.-H., 2003,
PASP, 115, 1006
\bibitem[Zhang Y.-X. \& Zhao Y.-H. 2004]{b33} Zhang Y.-X., \& Zhao Y.-H., 2004,
A\&A, 422, 1113

\end{thebibliography}
\end{document}